\documentclass[12pt]{iopart}
\usepackage{graphicx}
\usepackage{color}
\usepackage{iopams}

\def\m{\mu}
\def\n{\nu}

\def\x{\xi}

\newcommand{\be}{\begin{equation}}
\newcommand{\ee}{\end{equation}}

\newcommand{\ba}{\begin{eqnarray}}
\newcommand{\ea}{\end{eqnarray}}

\eqnobysec

\begin{document}


\begin{titlepage}

\title{
     String-like Lagrangians from a generalized geometry}
\author{S Morris}
\address{Department of Mathematical
Sciences, University of Liverpool, Liverpool, L69 3BX,
UK}

\begin{abstract}\noindent
This note will use Hitchin's generalized geometry and a model of
axionic gravity developed by Warren Siegel in the mid-nineties to
show that the construction of Lagrangians based on the inner
product arising from the pairing of a vector and its dual can lead
naturally to the low-energy Lagrangian of the bosonic string.

\end{abstract}
\maketitle

\thispagestyle{empty}

\end{titlepage}

\noindent String theory is one of the most popular theories of
quantum gravity. It is widely believed to predict gravity, while
remaining quantum mechanically finite. Nevertheless, unlike
relativity, or gauge theory, it cannot be derived from a simple
postulate. This note, inspired by Hitchin's generalized geometry,
and a model of axionic gravity developed by Warren Siegel in the
mid-nineties, will argue that the low energy Lagrangian of the
bosonic string follows naturally from the use of an inner product
based on the pairing between a vector and its dual, rather than
the, arguably less fundamental, standard Riemannian metric.

In the generalized geometries developed by Nigel Hitchin and his
students \cite{Hitchin:2004ut,gualtierithesis,Hitchin:2005in}
geometric objects that are normally defined solely on the tangent
bundle $T$, or cotangent bundle $T^*$, are redefined on the vector
bundle $T\oplus T^*$. These geometries are endowed with a natural
metric arising from the inner product between elements of a vector
space and its dual. This inner product has, on a d-dimensional
manifold, an $O(d,d)$ symmetry similar to that of the T-dualities
found in string theory. The choice of subspaces of $T\oplus T^*$
that are positive, or negative, definite with respect to the
natural metric breaks the symmetry to $O(d)\times O(d)$ and in
turn leads to the generation of a positive definite metric,
corresponding to the standard metric on a Riemannian manifold, and
a b-field \cite{gualtierithesis,Hitchin:2005in}.

A model of axionic gravity having similar properties was developed
by Warren Siegel in the mid-nineties \cite{Siegel:1993xq}. It was
formulated by adding a second vielbein to an Einstein-Cartan
theory of gravity, and has the Lagrangian of the closed, oriented
bosonic string at low energies. The extra vielbein combines with
the standard one to form an object transforming as a vector of
$O(d,d)$. This paper will reappraise Siegel's model in the light
of the recent development of generalized geometries. It will argue
that the vielbeins can be understood as a set of $d$ sections of
$T\oplus T^*$, coupling in the same way, and having the same
transformation properties and relationship to the metric and
b-field.

In addition to an $O(d,d)$ duality symmetry, the Lagrangian
describing Siegel's model has a $GL(d)$ gauge symmetry that leaves
the physical metric and b-field unchanged. The gauge potential for
this symmetry is a combination of the vielbeins rather than an
independent field. The Lagrangian also contains a scalar field,
corresponding to the dilaton, required to make the measure duality
invariant, and terms constructed from the vielbeins corresponding
to the Ricci scalar and $H^2$ terms from the low-energy action for
the bosonic string (the $H^2$ term accounts for the behaviour of
the b-field via the relation $H=\mathrm{d}b$). The Ricci scalar
can be regarded as a curvature term for the $GL(d)$ gauge
symmetry, and the $H^2$ term as the analogue of the
$F_{\m\n}F^{\m\n}$ term arising in conventional theories of vector
fields, but with the ``vectors" living in $T\oplus T^*$ rather
than $T$ alone. These are the terms one might expect, given the
model's field content, implying that the low energy Lagrangian of
the bosonic string follows naturally from the use of an inner
product based on the pairing of elements of the tangent and
cotangent spaces in place of an inner product based on a
Riemannian metric. This new inner product is arguably more
fundamental than the original one; it exists on any differential
manifold and does not require the existence of a Riemannian
metric.
\renewcommand{\thefootnote}{\alph{footnote}}
\section{Axionic Gravity and Generalized Geometry}\label{axionic}
In 1993 Warren Siegel proposed a model of axionic gravity inspired
by string theory and based, for a $d$ dimensional manifold, on
objects transforming as vectors of $O(d,d)$ \cite{Siegel:1993xq}.
He began with the observation that the action for bosonic string
theory can be written as

\begin{equation}
S=\int \partial_+X^m \partial_- X^n e_{mn}, \end{equation} where
$\partial_\pm$ are lightlike derivatives on the string's
worldsheet, and
\begin{equation}
e_{mn}=g_{mn}+b_{mn},\end{equation} $g$ and $b$ being the usual
metric and two-form respectively. He goes on to note that

\begin{equation}\label{e}
e_{mn}=e_{ma}e_n^a,\end{equation} where $e_n^a$ and $e_{ma}$  can
be thought of as right and left handed vielbeins. This expression
is invariant under $GL(d)$ transformations of the form
\cite{Siegel:1993xq}

\begin{equation}
e_{ma}\rightarrow  \Lambda_a ^b e_{mb},\label{GL1} \qquad
e_m^a\rightarrow  (\Lambda^{-1})^a_b e_m^b.\end{equation}

When constructing his theory, Siegel found it more useful to use
$e_{am}$ and the inverse of $e_m^a$, $e_a^m$ as fundamental
fields. They can be combined to form an object\footnote{The form
of $E_a$ given here differs slightly from that given in
\cite{Siegel:1993xq}; the order of the two vielbeins has been
reversed to make comparisons with generalized geometry easier.}

\begin{equation}
E_a=\left(\begin{array}{c} e_a^m \\ e_{ma}\end{array}\right)
\end{equation} that transforms as a vector of $O(d,d)$ (on the $m$
index), and the $GL(d)$ transformations given above (on the $a$
index).

It is also useful to define an indefinite metric that reflects the
$O(d,d)$ structure

\begin{equation}\label{sL}
L=\left(\begin{array}{cc} 0 & \mathbb{I}_d \\ \mathbb{I}_d & 0
\end{array}\right).\end{equation}

A similar metric is found in the generalized geometries recently
developed by Nigel Hitchin and his students (a good introduction
to generalized geometry, which I have followed here, can be found
in \cite{gualtierithesis}). Consider a vector space $V$ of
dimension $d$, and its dual $V^*$. It is possible to define a new
vector space $V\oplus V^*$, which is naturally endowed with an
inner product

\begin{equation}\label{innerproduct}
\langle X+\xi, Y+\eta \rangle = \frac{1}{2}\left(
\xi(Y)+\eta(X)\right),
\end{equation}
\noindent where $X,Y\in V$ and $\x,\eta\in V^*$. It defines a
metric $L$ on $V\oplus V^*$ and we can write, in a basis where
indices  ranging from 1 to $d$ lie in $V$ and those ranging from
$d+1$ to $2d$ lie in $V^*$,

\begin{equation}\label{L}
L=\left(\begin{array}{cc} 0 & \mathbb{I}_d \\ \mathbb{I}_d & 0
\end{array}\right) .
\end{equation}

\noindent

A generalized geometry takes $V$ to be the tangent space $T$ of a
manifold, and $V^*$ its dual, the cotangent space $T^*$. If the
identifications \begin{equation}\label{ggsgident} X^m_a=e^m_a,
\qquad \x_{ma}=e_{ma}\end{equation} are made, with $X_a$ being a
set of $d$ vectors and $\x$ being a set of $d$ one-forms, the
metric $L$ is the same as that in eq. \ref{sL}, the first hint
that Siegal's model can be understood in terms of a theory based
on sections of $T\oplus T^*$.

It is also possible to define a positive definite
metric\footnote{Actually, the metric $G$ considered by Siegel is
indefinite so that the standard metric $g$ derived from it can
incorporate time. Most of the work studying generalized geometry
up to now has considered a positive definite metric, a convention
I have followed here. The underlying mathematical structures are
basically the same except $O(d)\times O(d)$ becomes
$O(1,d-1)\times O(1,d-1)$.} $G=LM$, where
\begin{equation}\label{siegM} M=(LE_a)g^{ab}(LE_b)^T-L,
\end{equation} $g^{ab}$ being the inverse of \begin{equation}
g_{ab}=\frac{1}{2}E^T_aLE_b.\end{equation} This equation, which
differs from that in \cite{Siegel:1993xq} because I have used a
different definition of $E_a$, follows \cite{Siegel:1993xq} from
the fact that $M$ can be written in terms of vielbeins $V$ obeying
\begin{equation}
M=VV^T, \qquad L=VLV^T. \end{equation} The eigenvalues of
$\mathbb{I}_{2d}+L$ are 2 and 0, both of which have $d$
corresponding eigenvectors. Then

\begin{equation}\label{mgdelab}M_{mn}+L_{mn}=2\mathcal{E}_{ma}\mathcal{E}^T_{an},\end{equation} where $\mathcal{E}_{ma}$ is some $2d\times d$ matrix obeying $GL\mathcal{E}=L\mathcal{E}$. Equation \eref{siegM}
can then be derived using the $GL(d)$ symmetry \eref{GL1}, the
equation above corresponding to the case
$g_{ab}=\frac{1}{2}\delta_{ab}$.

Like the indefinite metric $L$, $M$ can be found in generalized
geometry. If $C_+$ is a positive definite (with respect to $L$)
subspace of $T\oplus T^*$, and $C_-$ its negative definite
orthogonal compliment, the positive definite metric
\cite{gualtierithesis,Hitchin:2005in} is

\begin{equation}
M=\langle\; , \; \rangle_{C_+}-\langle\; , \;
\rangle_{C_-}.\end{equation}

\noindent Using the indefinite metric $L$ to identify $T\oplus
T^*$ and its dual, it is possible to show that $G=LM$ can be
regarded as a map from $T\oplus T^*$ to itself. The  $+1$ ($-1$)
eigenspace of this map is $C_+$ ($C_-$), which enables an
 expression for $G$ in terms of symmetric and
antisymmetric tensors to be found, the origin of the metric and
$b$-field in generalized geometry. It is \cite{gualtierithesis}

\begin{equation}\label{G}
G=\left(\begin{array} {cc} -g^{-1}b & g^{-1} \\ g-bg^{-1}b
&bg^{-1}\end{array}\right) .\end{equation} It is also possible
\cite{gualtierithesis} to show that within $C_\pm$
\begin{equation}\label{posdefdef}\xi=(b\pm g)X.\end{equation} This is consistent with the identifications made in
\eref{ggsgident}, the eigenvectors of $\mathbb{I}_{2d}+L$ with
eigenvalue $2$ corresponding to the $+1$ eigenvectors of $G$.
Equation \eref{posdefdef} is then the same as \eref{e}.

In Siegel's model, the derivative $\partial_m$ is coupled to $E$
by defining \begin{equation}\label{oddderivative}
\partial_M=\left( \begin{array}{c} 0 \\ \partial_m \end{array}
\right), \quad e_a=E_a^M\partial_M=e^m_a\partial_m.\end{equation}
Derivatives transform as one-forms, so this is the coupling that
would be expected were the relationship between Siegel's model and
generalized geometry outlined above to hold.

It is also useful to define

\begin{equation}\label{fabc}
f_{abc} = \frac{1}{2} E^T_c L e_a E_b, \qquad f_{abcd} =
\frac{1}{2}(e_a E_b^T) L (e_c E_d). \end{equation}

\noindent $f_{ab}^{\;\;\;c}$ has the same transformation
properties as a $GL(d)$ connection, so it's possible
\cite{Siegel:1993xq} to associate the $GL(d)$ symmetry found in
Siegel's model with a covariant
derivative\begin{equation}\label{gldcodiv}
\nabla_a=e_a-f_{ab}^{\;\;\;c}. \end{equation}  The minus sign
ensures that the metric is covariantly constant. Note that a
theory with a $GL(d)$ gauge symmetry can be constructed from the
$E_a$'s alone; there's no need to introduce a separate $GL(d)$
gauge potential. To do so, one would need to replace $e_a$ with
the covariant derivative \eref{gldcodiv} in \eref{fabc}, leading
to \cite{Siegel:1993xq}
\begin{eqnarray}
F_{abc}&=&\frac{1}{2}E_c^TL\nabla_aE_b = 0 \\
\label{Fabcd}
F_{abcd}&=&\frac{1}{2}(\nabla_aE^T_b)L(\nabla_cE_d)=f_{abcd}-f_{ab}^{\;\;\;e}f_{cde}.\end{eqnarray}

Siegel relates his theory to a version of Cartan's theory of
gravity itself based upon a $GL(d)$ gauge theory. In that theory
the fundamental gravitational fields are a vielbein $e$ and an
independent tangent space metric, which is required to be
covariantly constant. It can be identified with the theory of
axionic gravity given above by taking $e^m_a$ to be the vielbein
while regarding $e_{am}$ as a kind of matter field. The tangent
space metric $g_{ab}$ can then be related to the standard one
$g_{mn}$ via \cite{Siegel:1993xq}\begin{equation}
g_{ab}=e_a^me_b^ng_{mn}.\end{equation}

The torsion and curvature tensors, $T_{ab}^{\;\;\;c}$ and
$R_{abc}^{\;\;\;\;d}$ can be found using
\begin{equation}
[\nabla_a,\nabla_b]=T^{\;\;\;c}_{ab}\nabla_c+R_{abc}^{\;\;\;\;d}G_d^c,\end{equation}
which, using the covariant derivative given in \eref{gldcodiv},
implies \cite{Siegel:1993xq}
\begin{equation}\label{R}
 R_{abcd}=-F_{[a|d|b]c}, \qquad
T_{ab}^{\;\;\;c}=c_{ab}^{\;\;\;c}-f_{[ab]}^{\;\;\;c},\end{equation}
where $c_{ab}^{\;\;\;c}$ is defined via the relationship
$[e_a,e_b]=c_{ab}^{\;\;\;c}e_c$.

$R$ can be thought of a $GL(d)$ field strength, something that can
be seen by defining \begin{equation}
\omega_{mab}=-\frac{1}{2}E^T_bL\partial_m E_a,
\end{equation} so that $\omega_{abc}=-f_{abc}=e^m_a\omega_{mbc}$, and regarding
$\omega_{ma}^{\;\;\;b}$ as a one-form valued in the Lie algebra of
$GL(d)$, a standard $GL(d)$ gauge field. Then
\begin{equation}
R_{abc}^{\;\;\;\;\;d}=e^m_ae^n_b R_{mnc}^{\;\;\;\;\;d},\end{equation} where

\begin{equation}
R_{mnc}^{\;\;\;\;\;d}=d\omega+\omega\wedge \omega\end{equation} is
the expected gauge field strength. It can also be shown \cite
{Siegel:1993xq} that\begin{equation}
H_{abc}=\frac{1}{2}e^m_ae^n_be^p_c\partial_{[m}b_{np]}=\frac{1}{2}c_{[abc]}-f_{[abc]}.\end{equation}

The low energy field-theory Lagrangian of the closed, oriented
bosonic string is
\cite{Fradkin:1984pq,Fradkin:1985ys,Callan:1985ia}
\begin{equation}\label{10dsgaction}
\mathcal{L}=\phi^2\left(\tilde{R}-\frac{1}{12}H^2\right)+4g^{mn}\partial_m\phi\partial_n\phi
,\end{equation} where $\tilde{R}$ is the torsion free version of
the curvature scalar and $\phi$ is the dilaton. It is possible to
show, after a series of algebraic manipulations detailed in
\cite{Siegel:1993xq}, that the corresponding action can be written
in terms of the fields given earlier.

\begin{equation}\label{seigelfinalaction}
S=\int d^dx \sqrt{g}\mathcal{L}=\int d^dx \;\;
4\left\{\left[\nabla \Phi+\frac{1}{2}
(1\cdot\overleftarrow{\nabla})\Phi\right]^2+\Phi^2\left(
F^a_{\;\;[a}{}^b_{\;\;b]}+F^{ab}_{\;\;\;[ab]}\right)\right\}.
\end{equation}
\noindent The field $\Phi=g^{1/4}\phi$ absorbs the measure, and
renders it duality invariant. $F_{[ab]cd}$ is an analogue of the
$F_{\m\n}F^{\m\n}$ term for the vector field $E_a$. $F_{[a|d|b]c}$
can similarly be regarded as a $GL(d)$ curvature term. A $GL(d)$
analogue of $F_{\m\n}F^{\m\n}$ would be fourth-order in
derivatives of $E_a$, but this problem can be avoided because the
fact that the metric $g_{ab}$ transforms under $GL(d)$ means that
a $GL(d)$ invariant scalar can be constructed from a single copy
of the curvature.

\section{Conclusions}\label{conclusions}
So, within the context of a generalized geometry in which objects
that would have been defined on a manifold's tangent, or cotangent
bundles, $T$ or $T^*$, are instead defined on their direct sum
$T\oplus T^*$, it is possible to construct a simple model,
involving only scalar and vector fields, whose Lagrangian is the
field theory Lagrangian of the closed, oriented bosonic string at
low energies. This model is based on one constructed by Warren
Siegel in the mid-nineties. In Siegel's model the metric and
b-field found in string theory are constructed from two
independent vielbeins which combine to form objects transforming
as vectors of $O(d,d)$. I have argued that these vielbeins can be
thought of as a set of sections of $T\oplus T^*$, which also
transforms naturally as a vector under $O(d,d)$ and has the same
relation to the metric and b-field. The model has a separate
$GL(d)$ symmetry, which can be realized locally without having to
introduce a separate $GL(d)$ gauge field because there exists a
combination of the vielbeins that transforms in the same way. The
model also contains a scalar field, a kind of dilaton, which is
introduced to compensate for the duality transformations of the
measure. The Lagrangian contains a curvature term for $GL(d)$,
which corresponds to the Ricci scalar in other gravitational
theories, and a term analogous to both the $F_{\m\n}F^{\m\n}$ term
found in standard theories of vector fields, but applied to
sections of $T\oplus T^*$ instead of $T$ alone, and the $H^2$ term
of the Lagrangian for the bosonic string. These terms are not
overly complex, on the contrary, they are the ones that would be
expected given the symmetries of the model. The coupling between
vectors and one-forms around which this geometry is formulated is
arguably more fundamental than the standard one between vectors
alone involving a metric, and so necessitating the introduction of
a new field, even though it may be less familiar.

\section{Acknowledgements}
This work was supported, in part, by the STFC.


\section*{References}
\begin{thebibliography}{77}
\bibitem{Hitchin:2004ut}
  Hitchin N 2003
  Generalized Calabi-Yau manifolds,
  {\it Quart.\ J.\ Math.\ Oxford Ser.}  {\bf 54}  281
  ({\it Preprint} math.dg/0209099)

\bibitem{gualtierithesis}
  Gualtieri M 2004
  Generalized complex geometry
{\it Preprint} math.dg/0401221



\bibitem{Hitchin:2005in}
  Hitchin N 2005
  Brackets, forms and invariant functionals
  {\it Preprint} math.dg/0508618








\bibitem{Siegel:1993xq}
 Siegel W 1993
  Two vierbein formalism for string inspired axionic gravity
  {\it Phys.\ Rev.\ D} {\bf 47} 5453
  ({\it Preprint} hep-th/9302036)




\bibitem{Fradkin:1984pq}
  Fradkin E S and Tseytlin A A 1985 Effective Field Theory From Quantized Strings
  {\it Phys.\ Lett.\  B} {\bf 158} 316

\bibitem{Fradkin:1985ys}
  Fradkin E S and Tseytlin A A 1985
  Quantum String Theory Effective Action
  {\it Nucl. Phys. B} {\bf 261} 1.

\bibitem{Callan:1985ia}
  Callan C G, Martinec E J, Perry M J and Friedan D 1985
  Strings In Background Fields
  {\it Nucl. Phys. B} {\bf 262} 593




\end{thebibliography}
\end{document}